\documentclass{ws-ijmpa}
\usepackage[super,compress]{cite}
\usepackage{graphicx}
\usepackage{color}
\UseRawInputEncoding
\begin{document}
\markboth{V. M. Mostepanenko \& G. L. Klimchitskaya}{Recent Solution to the Casimir Puzzle Awaits
Its Experimental Confirmation}

%%%%%%%%%%%%%%%%%%%%% Publisher's Area please ignore %%%%%%%%%%%%%%%
%
\catchline{}{}{}{}{}
%
%%%%%%%%%%%%%%%%%%%%%%%%%%%%%%%%%%%%%%%%%%%%%%%%%%%%%%%%%%%%%%%%%%%%

\title{\uppercase{Recent solution to the Casimir puzzle awaits its experimental
confirmation}}

\author{\uppercase{V. M. Mostepanenko${}^{1,2}$} \lowercase{and}
\uppercase{G. L. Klimchitskaya${}^{1,2}$}}
\address{${}^1$Central Astronomical Observatory at Pulkovo of the
Russian Academy of Sciences, Saint Petersburg,
196140, Russia\\
${}^2$Peter the Great Saint Petersburg
Polytechnic University, Saint Petersburg, 195251, Russia\\
 vmostepa@gmail.com, g.klimchitskaya@gmail.com
}

\maketitle

%\begin{history}
%\received{Day Month Year}
%\revised{Day Month Year}
%\end{history}

\begin{abstract}
The plausible resolution of the Casimir puzzle implying that the dissipative Drude
model is not applicable in the area of transverse electric evanescent waves is
discussed. Calculations show  that for the propagating waves, as well for the
evanescent waves with transverse magnetic polarization, the Drude model can be
used in calculations of the Casimir force by the Lifshitz theory with no
contradictions with the measurement data. The lateral component of magnetic
field of the magnetic dipole oscillating near a metallic surface is computed for
the parameters of experiment in preparation which is aimed to directly check the
validity of the Drude model in the area of transverse electric evanescent waves.
By comparing with the case of graphene, whose dielectric response is spatially
nonlocal and possesses the double pole at zero frequency, it is hypothesized that
the success of the dissipationless plasma model in this area is also caused by
the presence of a double pole.

\end{abstract}

\ccode{PACS numbers: 12.20.Fv, 12.20.Ds, 68.65.Pq}

%\tableofcontents

\section{Introduction}

In the course of the research work in physics, it happens that the measurement data
are in contradiction with the theoretical predictions. If this is the case, it is usual to
check whether there is some drawback in the experimental procedures and data analysis
or we face an inadequacy of the used theory. What's more, if the theory is of fundamental
character and the measurements were repeated by the independent groups in different
laboratories, the problem is aggravated.

Just this had happened with the Casimir effect, the macroscopic force of quantum origin
emerging between two bodies separated by a narrow gap.\cite{1,2,3} During the last
twenty years, it was undeniably demonstrated that precise measurements of the
Casimir force agree with theoretical predictions of the fundamental Lifshitz theory\cite{4}
using the dissipationless plasma model and exclude predictions of the dissipative Drude
model, which is generally accepted and confirmed otherwise.

It was also found that the Casimir entropy calculated within the Lifshitz theory for
metallic bodies with perfect crystal lattices using the Drude model violates the Nernst
heat theorem.\cite{5} The fact that the Nernst heat theorem is satisfied when the plasma
model is used only adds to the complexity in this situation because the plasma model
does not take into account the dissipation processes of conduction electrons which is
an actually existing effect. That is why the theory-experiment comparison in this field of
physics was often called ``the Casimir puzzle".\cite{6,7,8,9,10}

In this paper, we summarize recent achievements which allow to conclude that, from
the theoretical point of view, the Casimir puzzle has already found its solution. The
first of this achievements is the proof of the fact that the dissipative Drude model can be
safely used for calculation of the Casimir force with no contradiction with the
measurement data for the on-the-mass-shell (propagating) waves of any polarization
and for the transverse magnetic off-the-mass-shell (evanescent) waves. The conclusion
was made that in the region of the transverse electric evanescent waves the Drude model
describes the dissipation processes incorrectly.

The second achievement is the suggestion of a new experiment in the field of classical
electrodynamics which could confirm this conclusion. Below we present the
computational results for the experimental parameters, which demonstrate a
considerable difference of the predicted effect when using the Drude and plasma models
in the area of the transverse electric evanescent waves.

Finally, we compare the Casimir force between metallic plates and between two
graphene sheets whose electromagnetic response is found from the
first principles of quantum field theory. This leads to a conclusion
that an adequate dielectric function in the area of transverse electric evanescent waves
should be spatially nonlocal and (like the plasma model) possess the second-order pole
at zero frequency.

\section{Experimental situation}

During the last twenty years, precise experiments on measuring the Casimir force between
metallic surfaces were performed in the following five stages. In the first stage, the gradient of
the Casimir force has been measured in three successive experiments by Ricardo Decca
employing a micromechanical torsional oscillator.\cite{11,12,13} The predictions
of the Lifshitz theory used the extrapolations of the optical data\cite{15} to zero
frequency by means of the dielectric functions of the Drude and plasma models
\begin{equation}
\varepsilon_D(\omega)=1-\frac{\omega_p^2}{\omega(\omega+i\gamma)},\qquad
\varepsilon_p(\omega)=1-\frac{\omega_p^2}{\omega^2},
\label{eq1}
\end{equation}
\noindent
where $\omega_p$ is the plasma frequency and $\gamma$ is the relaxation parameter. As
a result, the predictions using an extrapolation by means of the dissipative Drude model were
excluded at the 99.9\% confidence level and the predictions using
the dissipationless plasma model were found consistent with the measurement data.

In the second stage, the gradient of the Casimir force has been measured by
Umar Mohideen between Au-Ni and Ni-Ni surfaces using an atomic force microscope.\cite{16,17,18}
An important new feature of these experiments is that for a magnetic metal Ni the theoretical
predictions using the Drude and plasma models switch places. Despite this, the predictions
of the Lifshitz theory using the Drude model were excluded and using the plasma model were
found consistent with the measurement data.

The third stage marked the conclusive rejection of theoretical predictions for the Casimir force using
the Drude model in the differential force measurement by Ricardo Decca.\cite{19} In
this seminal experiment, the predicted force values computed using the Drude and
plasma models differ by up to a factor of 1000. While the values obtained using the plasma model
were found consistent with the data, the values computed by means of the Drude model have
nothing to do with these data.

In the fourth stage of precise measurements of the Casimir interaction, Umar Mohideen has measured it
at separations up to more than 1~$\mu$m by means of an atomic force microscope,\cite{20,21,22}
whereas all the previous experiments were performed at separations up to a few hundred nanometers.
In doing so, the special measures for cleaning the interacting surfaces by means of the UV radiation
and Ar ions bombardment were undertaken. As to the results obtained, they were the same as before,
i.e., in favor of extrapolation by means of the plasma model.

Finally, at the fifth stage, Ricardo Decca performed the differential force measurement where the
predictions of the Lifshitz theory using the Drude model were excluded at separations up to 4.8~$\mu$m,
whereas the predictions of the same theory using the plasma model were consistent with the data.\cite{23}

For the sake of completeness, we also mention the single experiment which claimed an
agreement with theoretical predictions using the Drude model.\cite{24} It was demonstrated,
however, that this experiment was burdened with serious flaws by subtracting up to an order
of magnitude larger force, than the Casimir one, from the measurement data and by disregarding the
role of surface imperfections which are unavoidably present on spherical surfaces of the centimeter-size
radia.\cite{25,26}

It can be concluded that the measurement data exclude the predictions of the
Lifshitz theory using the dissipative Drude model and are consistent with those
using the dissipationless plasma model.
{{It should be noted that this conflict can be considered in a wider context than
between the Drude and plasma models. Although these models have a number of
experimental confirmations in certain domains of low and high frequencies,
respectively, both of them are of more or less phenomenological character and
should be treated as the approximate ones. The more general question is how
the electric susceptibility $\alpha(\omega)=\varepsilon(\omega)-1$ behaves
when the frequency goes to zero. If $\omega^2\alpha(\omega)$ goes to zero as
$\omega$, as it holds for the Drude model, we have a linear in temperature
correction to the Casimir pressure at short separations,\cite{2} which is in
contradiction with the measurement data. If, however, the quantity
$\omega^2\alpha(\omega)$ does not vanish when $\omega$ goes to zero, as it
holds for the plasma model, the big linear in temperature correction is
removed, and the theoretical predictions agree with the measurement data.

Thus, the experimental results exclude the class of the frequency-dependent
electric susceptibilities satisfying the condition
\begin{equation}
\lim_{\omega\to 0}\omega^2\alpha(\omega)=0.
\label{eq1a}
\end{equation}
}}
\noindent
Keeping in mind that dissipation of conduction electrons is the
physical effect which manifests itself in many electromagnetic phenomena, this conclusion should
not be understood in a sense that the above mentioned experiments testify the fallacy of the Drude model
and the correctness of the plasma one. The problem is in fact more intricate
{{and calls for a careful analysis of different relationships between the frequency and
the wave vector as well as for the proper allowance for the spatial dispersion
(see Secs.~3 and 5).
}}

\section{Contribution of Dissipation to the Casimir Pressure for Propagating and Evanescent Waves}

Using the Lifshitz formula written along the real frequency axis, the Casimir pressure between two
metallic plates spaced at a distance $a$ at temperature $T$ can be presented as the sum of four
contributions
\begin{equation}
P(a,T)=P_{\rm TM}^{\,\rm prop}(a,T)+P_{\rm TE}^{\,\rm prop}(a,T)+
P_{\rm TM}^{\,\rm evan}(a,T)+P_{\rm TE}^{\,\rm evan}(a,T),
\label{eq3}
\end{equation}
\noindent
where
\begin{eqnarray}
&&
P_{\rm TM,TE}^{\,\rm prop}(a,T)=-\frac{\hbar}{2\pi^2}\int\limits_{0}^{\infty}\!\!d\omega
\coth\frac{\hbar\omega}{2k_BT}
\int\limits_0^{\omega/c}\!\!kdk {\rm Im}\left[
\frac{qr_{\rm TM,TE}^2(\omega,k) e^{-2aq}}{1-r_{\rm TM,TE}^2(\omega,k) e^{-2aq}}\right],
\nonumber\\
&&
P_{\rm TM,TE}^{\,\rm evan}(a,T)=-\frac{\hbar}{2\pi^2}\int\limits_{0}^{\infty}\!\!d\omega
\coth\frac{\hbar\omega}{2k_BT}
\int\limits_{\omega/c}^{\infty}\!\!kdk q {\rm Im}
\frac{r_{\rm TM,TE}^2(\omega,k) e^{-2aq}}{1-r_{\rm TM,TE}^2(\omega,k) e^{-2aq}}.
\label{eq4}
\end{eqnarray}
\noindent
Here, $k_B$ is the Boltzmann constant, $\omega$ is the frequency, $k$ is the magnitude of the wave
vector projection on the plane of plates, $r_{\rm TM,TE}$ are the reflection coefficients for the waves with
transverse magnetic (TM) and transverse electric (TE) polarizations, and $q=(k^2-\omega^2/c^2)^{1/2}$.

The quantities $P_{\rm TM,TE}^{\rm prop}$ are the contributions of the on-the-mass-shell
(propagating) waves with $k\leq \omega/c$ to the Casimir pressure and $P_{\rm TM,TE}^{\rm evan}$ --
of the off-the-mass-shell (evanescent) waves for which  $k> \omega/c$. Recently, all four contributions
to (\ref{eq3}) were computed for two Au plates\cite{27} using the Drude and plasma models (\ref{eq1})
at separations exceeding 0.5~$\mu$m, where the contribution of core electrons to
the Casimir pressure is negligibly small.

It was found that the contributions of the TM polarization to the Casimir pressure computed using
the Drude and plasma models
\begin{equation}
P_{\rm TM,D}\equiv P_{\rm TM,D}^{\rm prop} + P_{\rm TM,D}^{\rm evan},\quad
P_{\rm TM,p}\equiv P_{\rm TM,p}^{\rm prop} + P_{\rm TM,p}^{\rm evan},
\label{eq5}
\end{equation}
\noindent
are nearly equal, $P_{\rm TM,D}\approx P_{\rm TM,p}$, although their constituents are different. Thus,
$P_{\rm TM,p}^{\rm evan}=0$ because $q$ is real for the evanescent waves and, as a result,
$P_{\rm TM,p}= P_{\rm TM,p}^{\rm prop}$. As to the Drude model, the difference between
$P_{\rm TM,D}^{\rm prop}$ and $P_{\rm TM,p}$ is compensated by $P_{\rm TM,D}^{\rm evan}$
which is of the opposite sign. Thus, one can compute the contribution of the TM polarization to the
Casimir pressure using the dissipative Drude model without contradiction with the measurement data.

For the TE polarization, one can write similar to (\ref{eq5}) definitions, but the situation is more
complicated. If the plasma model is used in computations, we find that
$P_{\rm TE,p}= P_{\rm TE,p}^{\rm prop}$ because $P_{\rm TE,p}^{\rm evan}=0$ due to the fact that
$q$ is real. The contribution $P_{\rm TE,D}^{\rm prop}$ deviates moderately from $P_{\rm TE,p}$.
This deviation is in the limits of the measurement errors and does not influence the theory-experiment
comparison. However, the contribution $P_{\rm TE,D}^{\rm evan}$, which is of the opposite sign,
not only compensates the difference between $P_{\rm TE,D}^{\rm prop}$ and $P_{\rm TE,p}$, but
makes the contribution $P_{\rm TE,D}\equiv P_{\rm TE,D}^{\rm prop} + P_{\rm TE,D}^{\rm evan}$
significantly different from $P_{\rm TE,p}$. Just this difference is observed experimentally and results in an
exclusion of the Drude model by the measurement data.

The conclusion is that the contribution $P_{\rm TE}^{\rm prop}$ can also be calculated using the Drude
model with no contradiction with the measurement data. The contradiction is caused by application of the
Drude model only in the area of TE polarized off-the-mass-shell (evanescent) waves. The question
on whether the Drude model has sufficient experimental confirmation on this region is considered in the
next section.

\section{Independent Test of the Drude Model for Transverse Electric Evanescent Waves}

As was mentioned in Sec. 1, the Drude model has numerous experimental confirmations. The most
of them, however, falls in the area of on-the-mass-shell (propagating) waves with any polarization.
There are only a few exceptions. Thus, the electromagnetic response of metals and associated
dielectric functions in the area of TM polarized evanescent waves can be probed with sufficient
precision by means of the near field optical microscopy\cite{28,29} and by the methods developed
in physics of surface plasmon polaritons.\cite{30} As to the area of TE polarized evanescent waves, here
the dielectric properties of metals can be probed by using the techniques of total internal reflection
and frustrated total internal reflection.\cite{31} These methods, however, are applicable for $k$ only
slightly exceeding $\omega/c$ and, thus, are incapable to test the validity of the Drude model in
the area required for calculation of the Casimir force. Therefore, at the moment, the Drude model is
lacking of sufficient experimental confirmation in the area of TE polarized evanescent waves.

Recently, the alternative test of the Drude model in this area was proposed.\cite{32,33} It was shown
that the lateral component of magnetic field of small oscillating magnetic dipole
${\bf m}=(0,0,m_0{\rm exp} (-i\omega t))$ placed at the height $h$ above a metallic surface measured
at the same height $h$ at the separation $x$ along the $x$-axis directed from the dipole to the
observation point is given by
\begin{equation}
B_x(\omega,x)=m_0\int\limits_{\omega/c}^{\infty}\!\!k^2dkJ_1(kx)r_{\rm TE}(\omega,k){\rm e}^{-2hq},
\label{eq6}
\end{equation}
\noindent
where $J_1$ is the Bessel function. In doing so, the electric field produced by this dipole is smaller than
the magnetic one by many orders of magnitude.

As is seen from (\ref{eq6}), the lateral component $B_x$ is fully determined by the TE polarized
evanescent waves. Thus, by measuring it and comparing the measurement results with computations by
 (\ref{eq6}) using the Drude model, one can check whether this model is applicable.\cite{32,33}
Here, we perform numerical computations of the lateral component of magnetic field  (\ref{eq6})
for the parameters of experiment by Umar Mohideen in preparation, i.e., for
$m_0=2.776\times 10^{-3}$~Am$^2$, $h=3$~cm and two values of the oscillation frequency
$\omega=(2\pi\times 15)$~rad/s and $\omega=(2\pi\times 25)$~rad/s.

%%%%%%%%%%%%%%%__Figure_1__%%%%%%%%%%%%%%%%%%%%%%%%%%%%%%%%%
\begin{figure}[t]
\vspace*{-10.5cm}
\hspace*{-2.5mm}
\centerline{\includegraphics[width=16.5cm]{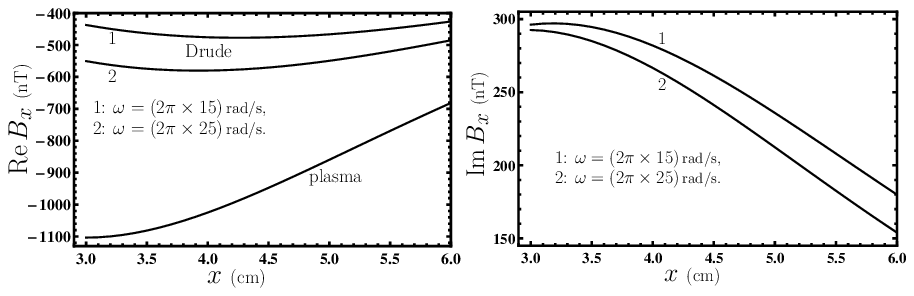}}
\vspace*{-9.2cm}
\caption{Computational results for the (left) real and (right) imaginary parts of the lateral
 component of magnetic field of the magnetic dipole oscillating near a copper plate are presented as
 the functions of separation. Computations are performed by means of the Drude and plasma models
 for different oscillation frequencies (see the text for further discussion).
 \label{f1}}
\end{figure}
%%%%%%%%%%%%%%%%%%%%%%%%%%%%%%%%%%%%%%
In Fig.~1 (left) the computational results for the real part of the lateral component of magnetic field
are presented as the function of separation by the lines 1 and 2 for the dipole frequencies
indicated above if the Drude model is used and by the bottom line if the plasma
model is used in computations. As the plate metal, copper was chosen for which the Drude model
parameters are\cite{34} $\omega_p=1.12\times 10^{16}$~rad/s and
$\gamma=1.38\times 10^{13}$~rad/s. In Fig.~1 (right), similar results are presented for the imaginary
part of the lateral component of magnetic field computed using the Drude model (if the plasma model
is used, $r_{\rm TE}$ is real and ${\rm Im}B_x=0$).

As is seen in Fig.~1 (left), the theoretical predictions of the Drude model given by line 2 and plasma model
differ by up to a factor of 2. This difference increases with decreasing $\omega$ and decreases with
increasing $x$. Although it should not be expected that the results of this experiment will be found in
agreement with the predictions obtained using the dissipationless plasma model, they may
demonstrate an incorrectness of the Drude model in the area of TE polarized evanescent waves and supply
us with the experimentally valid real and imaginary parts of $B_x$.

\section{Lessons of Graphene}

The distinctive feature of graphene, the two-dimensional sheet of carbon atoms, is that at
energies below 3~eV characteristic for the Casimir effect it can be described as a
collection of massless or light quasiparticles satisfying Dirac equation with the Fermi velocity $v_F$
playing the role of the speed of light.\cite{35} Due to this, the dielectric response of graphene
to the electromagnetic field can be expressed via the polarization tensor $\Pi_{\alpha\beta}(\omega,k)$
with $\alpha,\beta=0,1,2...$, which is calculated exactly starting from the first principles of thermal
quantum field theory in (2+1) dimensions.\cite{36,37,38,39}

As a result, the spatially nonlocal longitudinal and transverse dielectric permittivities of graphene are
expressed via the polarization tensor\cite{3,40,41}
\begin{equation}
\varepsilon^{\rm L}(\omega,k)=1+\frac{1}{2\hbar k}\Pi_{00}(\omega,k), \quad
\varepsilon^{\rm Tr}(\omega,k)=1-\frac{c^2}{2\hbar k\omega^2}\Pi(\omega,k),
\label{eq7}
\end{equation}
\noindent
where $\Pi=k^2\Pi_{\mu}^{\mu}-q^2\Pi_{00}$.

The reflection coefficients on a graphene sheet and a graphene-coated plate are also expressed via the
permittivities (\ref{eq7}) and the permittivity of a substrate.\cite{36,37,42,43} It is
remarkable that already at zero temperature the transverse permittivity of graphene has
the double pole at $\omega=0$ as does the plasma model\cite{44}
\begin{equation}
\varepsilon^{\rm Tr}(\omega,k)=\left\{
\begin{array}{ll}
1-\frac{\pi\alpha kc}{2\omega^2}\sqrt{v_F^2k^2-\omega^2}, & |\omega|<v_Fk, \\
i\frac{\pi\alpha kc}{2\omega^2}{\rm sign}\omega\sqrt{\omega^2-v_F^2k^2}, & |\omega|>v_Fk,
\end{array}
\right.
\label{eq8}
\end{equation}
\noindent
where $\alpha$ is the fine structure constant.

It was proven\cite{44a} that the polarization tensor of graphene is defined uniquely and cannot be
subjected to any additional regularization with no violation of first physical principles. Thus,
the presence of the double pole in the spatially nonlocal contribution to $\varepsilon^{\rm Tr}$ of
graphene is the consequence of fundamental physics. Taking into account that
the dielectric permittivity (\ref{eq8}) has both the real and imaginary parts, it takes dissipation into
account as opposed to the plasma model.

Experiments on measuring the Casimir force in graphene systems performed by Umar Mohideen were
found in good agreement with theoretical predictions of the Lifshitz theory using the reflection
coefficients in terms of the polarization tensor.\cite{42,45,46,47} This suggests that the
correct permittivity of metals in the area of TE polarized evanescent waves should also be
spatially nonlocal and its transverse part should possess the double pole at zero frequency. In this
sense, a successful use of the plasma model in the theory-experiment comparison might be due to the
presence of the second-order pole.

Although an exact calculation of the polarization tensor for real metal is not yet possible, the
phenomenological candidate for its permittivity was proposed\cite{48,49}
\begin{equation}
\varepsilon_{\rm ph}^{\rm Tr}(\omega,k)=1-\frac{\omega_p^2}{\omega(\omega+i\gamma)}\left(1+i
\frac{vk}{\omega}\right),
\label{eq9}
\end{equation}
\noindent
where for the three-dimensional bodies $k=(k_1^2+k_2^2+k_3^2)^{1/2}$ and $v$ is the velocity
parameter of the order of $v_F$.

For the on-the-mass-shell (propagating) waves, it holds $ck\leq \omega$ and
$vk/\omega \sim (v_F/c)(ck/\omega)\leq v_F/c\ll 1$. As a result,
$\varepsilon_{\rm ph}^{\rm Tr}\approx \varepsilon_D$. It was shown that the predictions of the
Lifshitz theory using the permittivity (\ref{eq9}) are in good agreement with all experiments on
measuring the Casimir force\cite{49} and with the Nernst heat theorem.\cite{50} This permittivity
satisfies the Kramers-Kronig relations.\cite{49} It can be, however, applied only under a condition
$k< \gamma/v$ which assures that ${\rm Im}\varepsilon_{\rm ph}^{\rm Tr}>0$.

It should be stressed that the permittivity (\ref{eq8}) does not claim to be valid
in the strict sense. It is presented only as an example of the spatially nonlocal
response function which, on the one hand, almost coincides with the dissipative
Drude model for the on-the-mass-shell waves and, on the other hand, brings the
theoretical predictions in
agreement with the measurement data of all available measurements of the Casimir
force presumably due to the presence of a double pole in the wave vector-dependent
part. This is a matter for the future to establish the spatially nonlocal response
function of metals valid in the area of the off-the-mass-shell waves.

\section{Conclusions and Discussion}

In this paper, we have proposed the plausible solution to the Casimir puzzle suggesting that the Drude model
is invalid in the area of transverse electric evanescent waves where it lacks of sufficient experimental
confirmation. Otherwise, the dissipative Drude model can be safely used for calculation of the Casimir force.
It was shown that an agreement of theoretical predictions using the plasma model with the measurement data
is due to the occasional cancelations of the contributions of dissipation for the propagating waves of both
polarizations and for the TM evanescent waves.  As to the area of TE evanescent waves, the lesson of graphene
 suggests that there the success of the plasma model is not due to its dissipationless character but because of
the presence of a double pole at zero frequency in the dielectric function.

The suggested solution can be confirmed experimentally by measuring the lateral component of magnetic
field produced by the magnetic dipole oscillating in the proximity of a metal plate. This experiment in the
field of classical electrodynamics should give a clear evidence on whether or not the Drude model describes
correctly the electromagnetic response of metals to the off-the-mass-shell TE polarized waves. In this
paper, calculation of the lateral component of dipole magnetic field performed for the experimental parameters
shows a big difference between the theoretical predictions obtained using the Drude and plasma models.

In view of the above, we may hope that the final solution to the Casimir puzzle is at hand.

\section*{Acknowledgments}

The authors are grateful to Umar Mohideen for information on the parameters of his
experiment in preparation.
This work was supported by the
State Assignment for basic research (project FSEG--2023--0016).

\end{document}